\documentstyle[twoside,fleqn,espcrc2x,epsf]{article}

\newcommand{\AmS}{{\protect\the\textfont2
  A\kern-.1667em\lower.5ex\hbox{M}\kern-.125emS}}
\def\makeadmark#1{\hbox{$^{\rm #1}$}}
\def\simlt{\hbox{ \rlap{\raise 0.425ex\hbox{$<$}}\lower 0.65ex\hbox{$\sim$} }}
\def\simgt{\hbox{ \rlap{\raise 0.425ex\hbox{$>$}}\lower 0.65ex\hbox{$\sim$} }}
\def\that{{\hat t}}
\def\umin{u_{\rm min}}

\def\msun{ {\rm M_\odot} }

\def\apj{{\rm ApJ}}
\def\apjl{{\rm ApJL}}

\def\annrev{{\rm Ann.~Rev.~Astron.~Astrophys.}}
\def\etal{{\it et al.}}
\def\VEV#1{\left\langle #1\right\rangle}

\title{A Binary Lensing Event Toward the LMC: Observations and Dark Matter
   Implications}

\author{D.P.Bennett\rlap,\makeadmark{a,b,c,d}
C.Alcock\rlap,\makeadmark{a,b}
R.A.Allsman\rlap,\makeadmark{e}
D.Alves\rlap,\makeadmark{b,c}
T.S.Axelrod\rlap,\makeadmark{b,f}
A.Becker\rlap,\makeadmark{g}
K.H.Cook\rlap,\makeadmark{b}
K.C.Freeman\rlap,\makeadmark{f}
K.Griest\rlap,\makeadmark{a,h}
J.Guern\rlap,\makeadmark{a,h}
M.J.Lehner\rlap,\makeadmark{a,h}
S.L.Marshall\rlap,\makeadmark{a,b}
D.Minniti\rlap,\makeadmark{b}
B.A.Peterson\rlap,\makeadmark{f}
M.R.Pratt\rlap,\makeadmark{b,g}
P.J.Quinn\rlap,\makeadmark{i}
S.H.Rhie\rlap,\makeadmark{a,b}
A.W.Rodgers\rlap,\makeadmark{f}
C.W.Stubbs\rlap,\makeadmark{a,h}
W.Sutherland\rlap,\makeadmark{j} and
D.Welch\makeadmark{k}
\address{Center for Particle Astrophysics, University of California,
	Berkeley, CA 94720}
\address{Lawrence Livermore National Laboratory, Livermore, CA 94550}
\address{Department of Physics, University of California, Davis, CA 95616}
\address{Department of Physics, University of Notre Dame, Notre Dame, IN 46556}
\address{Supercomputing Facility, Australian National Univ.,
    Canberra, ACT 0200, Australia}
\address{Mount Stromlo and Siding Springs Obs.,
    Australian National Univ., Weston, ACT 2611, Australia}
\address{Departments of Physics and Astronomy, University of
    Washington, Seattle, WA 98195}
\address{Department of Physics, University of
	California San Diego, La Jolla, CA 92093-0350}
\address{European Southern Observatory, Garching, Germany}
\address{Department of Physics, University of Oxford,
    Oxford OX1 3RH, U.K.}
\address{Departments of Physics and Astronomy, McMaster Univ.,
    Hamilton, Ont., Canada L8S 4M1}
\thanks{Work performed at LLNL is
supported by the DOE under contract W-7405-ENG.  Work performed by the
CfPA personnel is supported by the NSF
through AST 9120005.  The work at MSSSO is supported by the Australian
Dept. of Industry, Science, and Technology.,  
K.G. acknowledges support from DOE OJI, Sloan, and Cotrell awards.  
C.S. acknowledges the support of the Packard and Sloan Foundations.}
}

\begin{document}

\begin{abstract}
The MACHO collaboration has recently analyzed 2.1 years of 
photometric data for about
8.5 million stars in the Large Magellanic Cloud (LMC). This analysis has
revealed 8 candidate microlensing events and a total microlensing optical depth
of $\tau_{\rm meas} = 2.9 {+1.4\atop -0.9}\times 10^{-7}$. This significantly
exceeds the number of events (1.1) and the microlensing optical depth predicted
from known stellar populations: $\tau_{\rm back} = 5.4\times 10^{-8}$, but it
is consistent with models in which about half of the standard dark halo mass is
composed of Machos of mass $\sim 0.5 \msun$. One of these 8 events appears to
be a binary lensing event with a caustic crossing that is partially resolved,
and the measured caustic crossing time allows us to estimate the distance to
the lenses. Under the assumption that the source star is a single star and not
a short period binary, we show that the lensing objects are very likely to
reside in the LMC. However, if we assume that the optical depth for LMC-LMC
lensing is large enough to account for our entire lensing signal, then the
binary event does not appear to be consistent with lensing of a single LMC
source star by a binary residing in the LMC. Thus, while the binary lens may
indeed reside in the LMC, there is no indication that most of the lenses
reside in the LMC.
\end{abstract}

\maketitle

\section{INTRODUCTION}
The MACHO project is a dedicated search
for Galactic dark matter in the form of MAssive
Compact Halo Objects (or Machos) using a gravitational microlensing
technique \cite{paczynski86,petrou}. By photometrically monitoring millions
of stars in the LMC, we are able to detect rare instances of gravitational
microlensing. Microlensing occurs when a dark object, such as a faint star 
or a Macho, comes close enough to the line of sight to one of our target 
stars in the LMC so that the gravitational field of the dark object magnifies 
the star by a significant amount (typically a factor of 2 or more).

The MACHO team has recently completed analysis of data for 8.5 million
stars spanning 2.1 years of observations \cite{pratt96,lmc2}. 
The analysis of this data has
revealed 8 candidate microlensing events with timescales in the range
$34\,{\rm days} < \that < 145\,{\rm days}$. The probability that a single
target star will be magnified by more than a factor of 1.34 is commonly
referred to as the microlensing optical depth, $\tau$. $\tau$ is a particularly
useful measure of the ``amount" of microlensing because it is
proportional to the
total mass in lensing objects and independent of the masses and velocities
of the lenses. (It does depend on the spacial distribution of the lenses,
however.) The MACHO data gives a microlensing optical depth in events
with an Einstein diameter crossing time $\that < 200\,{\rm days}$ of
$\tau_{\rm meas} = 2.9 {+1.4\atop -0.9}\times 10^{-7}$. This can be compared
to the predicted microlensing background due to lensing by known stellar
populations of $\tau_{\rm back} = 5.4\times 10^{-8}$ and the microlensing
optical depth of a standard halo model consisting entirely of Machos:
$\tau_{\rm halo} = 4.7\times 10^{-7}$. This suggests that perhaps half of the
mass of the ``standard" dark halo of our Galaxy is composed of Machos which
give rise to lensing events with $\that \simlt 200\,$days. The observed 
timescales imply a typical Macho mass of $\sim 0.5\msun$, but this is somewhat
model dependent. The simplest explanation of these microlensing results is
that a significant fraction of the Galactic halo is composed of white dwarfs.
The remainder of the dark halo could be in more massive Machos which give rise
to longer events, or it could be in some form other than Machos.
For some ``non-standard" halo models it is even possible that the halo
is composed entirely of Machos of mass $< 1 \msun$.

One difficulty with the microlensing dark matter search technique is that
the distance and mass of the lens cannot generally be determined on 
on an event-by-event basis. For most events, all information on the
distance, velocity, and mass of the lens is folded into a single measurable
parameter, the event timescale $\that$. This is why the lensing by known
populations of stars in the Galactic disk and the LMC cannot usually be
distinguished from lensing by halo objects on an event-by-event basis. 
Thus, the main argument indicating
that there is a previously undiscovered population of Machos in the halo
is the excess observed number of events and optical depth over the predicted
lensing background.


For certain types of exotic microlensing events additional parameters
can be measured that can help lift the degeneracy. One example
is a parallax event in which the lightcurve shows an asymmetry due to the
orbital motion of the Earth \cite{macho-parallax}. A parallax event lightcurve
fit provides an estimate of the transverse velocity of the lens projected to
the position of the solar system. We can then use our knowledge of the
velocity distributions in our Galaxy to make a lens distance estimate
good to about 30\%. Single \cite{alert9530} or 
binary \cite{bennett95,causticIAUC} lensing events in which a caustic
crossing occurs can also be used to estimate the distance to the lens.
The intrinsic lightcurve shape at a caustic crossing is so highly peaked
that the actual width of the caustic crossing lightcurve is determined
by the angular size of the source star. The observed caustic crossing time
can be compared to the known angular size of the source star to obtain
a projected transverse velocity, and this can be used to make a distance
estimate.

\begin{figure}[thbp]
\begin{center}
\leavevmode
\hbox{%
\epsfxsize=7.9cm
\epsffile{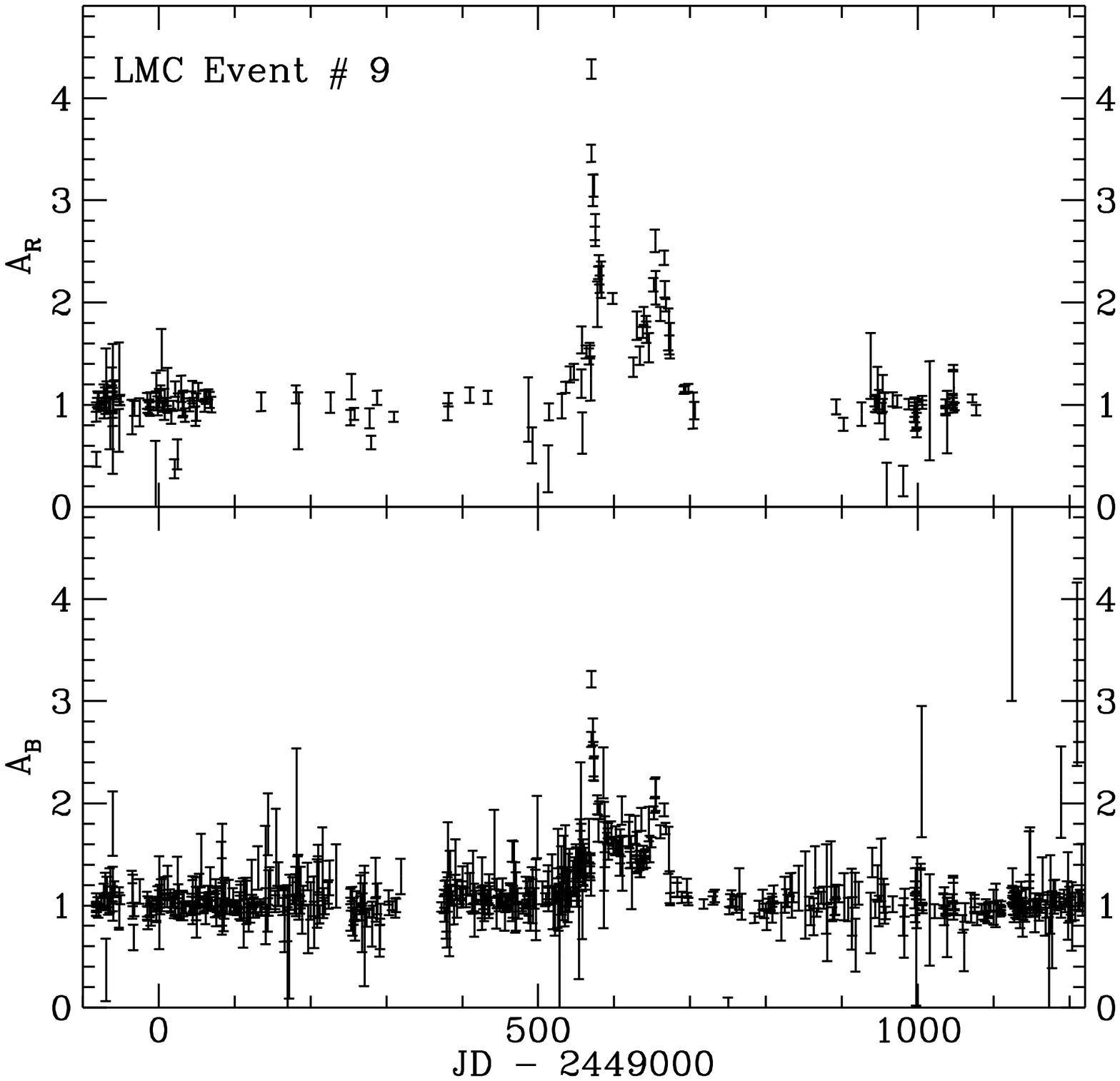}}
\vspace{-0.6cm}
\hbox{%
\epsfxsize=7.9cm
\epsffile{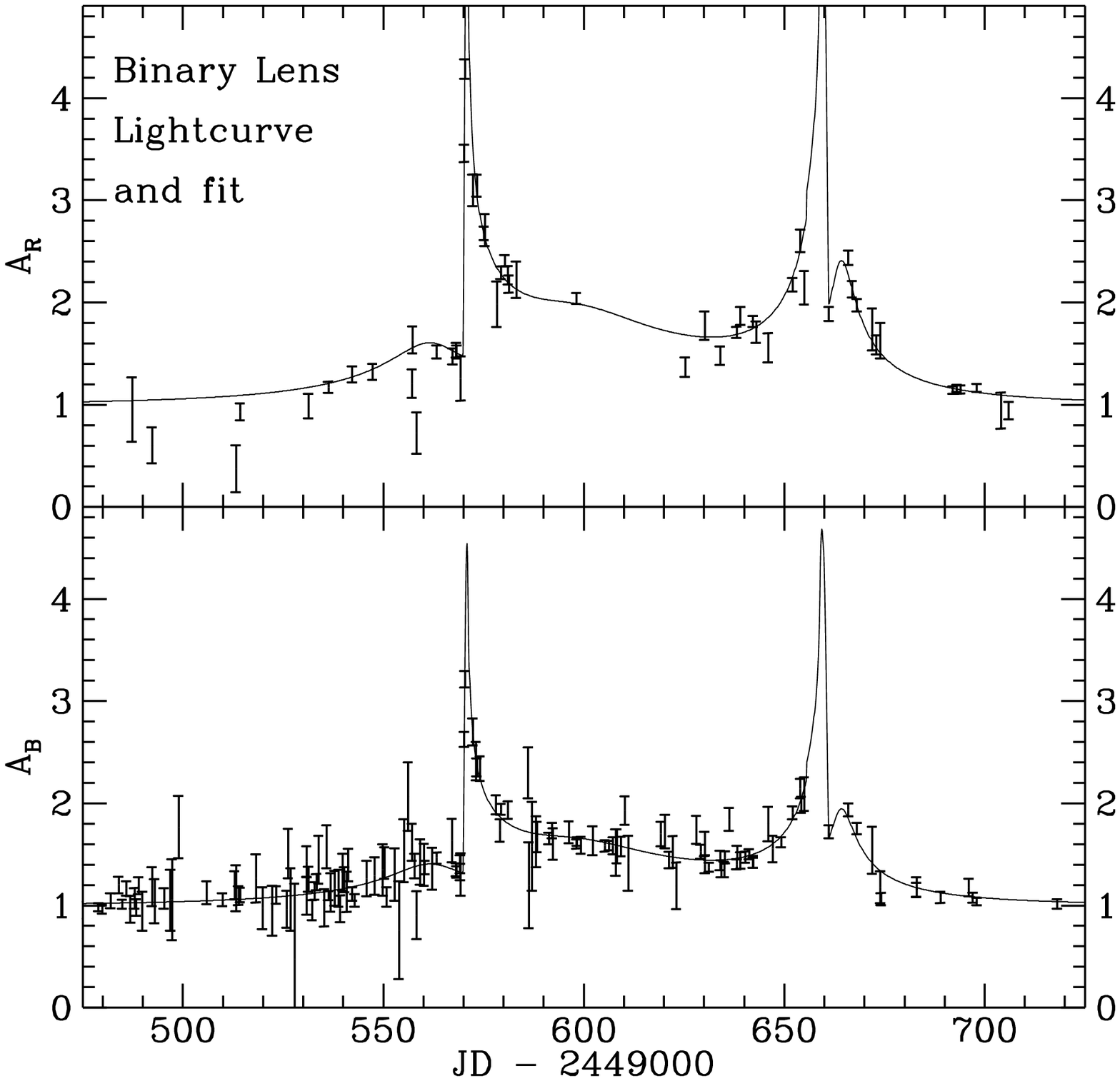}}
\end{center}
\vspace{-1.2cm}
\caption{The full dual color lightcurve of the LMC binary lensing event is
plotted in the upper panel while the lower panel shows 
a closeup of the lightcurve along with the fit lightcurve described in the
text. Many of the measurements in the red band have been contaminated
by a previously undetected CCD trap. A cut has been applied to remove
the measurements in which this contamination is expected to be severe.}
\label{fig_lmc_binary}
\end{figure}

\section{LMC Binary Event}

Figure~\ref{fig_lmc_binary} shows the lightcurve of the
LMC binary event as well as a closeup of the lightcurve along with the best fit
binary lens lightcurve.
There are many more data points in our blue band than in our red band
for two reasons. First, the Mt.~Stromlo 50" telescope has a equatorial
German mount that allows us to observe from both sides of the Pier. For
east of pier observations, this star falls on a non-functional region of
one of our CCD's in the red focal plane. In addition, for the west of
pier observations this star often falls on or near a previously undetected
CCD trap which smears the flux along a CCD column. We have removed all
red data in which this bad column comes within 6 pixels and 1.5 times
the PSF FWHM of the source star from the plot shown in 
Figure~\ref{fig_lmc_binary}. Due to the extreme crowding of stellar images
in this field, however, it is possible that this procedure does not 
remove all the red observations which might be contaminated.

A dual color binary microlensing lightcurve is described by 11 parameters.
Two parameters, $f_{0R}$ and $f_{0B}$, describe the unlensed brightness of 
the lensed star in each color band, and in the crowded fields in which
we work, we also need two parameters ($f_{uR}$ and $f_{uB}$) to describe 
the flux of unlensed stars that may be superimposed on the same resolving
element as the lensed star. Three of the intrinsic microlensing parameters are 
the same as the parameters for a single lens. These are: the Einstein 
diameter crossing time for the total mass, $\that$, the time of closest
approach, $t_0$, between the angular positions of the lens center of mass 
and the source star, and the distance of closest approach, $\umin$, which
is measured in units of the Einstein radius. There are three other intrinsic
microlensing parameters which are unique to the binary lens: $a$, the 
separation of the two lens masses in units of the Einstein radius, the
angle, $\theta$, 
between the lens axis and the apparent motion of the source in the
lens plane, and $\epsilon_1$, the mass fraction of mass \#1.
($\epsilon_2 \equiv 1-\epsilon_1$.) Finally, the parameter which can allow
us to learn more about the location of the lens is the time, $t_{\rm star}$,
it takes for the source star to move by one stellar radius with respect to the
angular position of the lens center of mass.

The formula for the lightcurve in terms of these parameters is well known
\cite{schneid,bennettrhie} but is rather complicated. We obtain the following
fit parameters: 
$f_{0R} = 0.259\pm 0.002$, $f_{0B} = 0.174\pm 0.001$, 
$f_{uR} = 0.741\pm 0.004$, $f_{uB} = 0.826\pm 0.002$,
$\that = 143.4\pm 0.2\,$days, $t_0 = 603.04\pm 0.02\,$days, 
$\umin = -0.055\pm 0.001$, $a = 1.6545 \pm 0.0008$, $\theta = 0.086\pm 0.001$,
$\epsilon_1 = 0.620\pm 0.002$, and $t_{\rm star} = 0.65\pm 0.18\,$days.
The binary lens fit gives $\chi^2=1489$ for 848 degrees of freedom or
a reduced $\chi^2$ of 1.76. (The best single lens fit gives a reduced $\chi^2$
of 6.03.)
The flux parameters have been normalized to give a total unlensed flux
of 1 in each color band, and the unmagnified magnitudes of the lensed
component are $V = 21.10\pm 0.10$ and $R = 20.76\pm 0.10$. 
Note that we must {\it assume}
that the source star is a single star and not a close binary because we
have only two observations on the caustic crossing feature. We have recently
demonstrated the capability to predict caustic crossing events
\cite{causticIAUC}, so future
caustic crossings can be observed much more frequently.

\section{Implications}

It is the parameter $t_{\rm star}$ that can potentially teach
us more about the lensing event because it depends on the finite size of the
source star. Using the magnitudes of the lensed star, and assuming between
0.2 and 0.6 magnitudes of extinction in $V$ and an LMC distance
modulus of 18.5, we estimate $M_V = 2.2\pm 0.2$. This implies that the
star is an A7-8 main sequence star with a radius 
of $R_{\rm star} = 1.5\pm 0.2\,R_\odot$.
Combining this with the observed $t_{\rm star}$ value yields
\begin{equation}
v_{\rm proj} = {R_{\rm star}\over t_{\rm star}} = 19\pm 6\,{\rm km/s} \qquad
  v_{\rm proj} \equiv {v_\perp\over x} ,
\label{eq_vproj}
\end{equation}
where $v_\perp$ is the transverse velocity of the lens with respect to the
line of sight and $x$ is the ratio of the lens and source distances.

\begin{figure}[ht]
\begin{center}
\leavevmode
\hbox{%
\epsfxsize=7.5cm
\epsffile{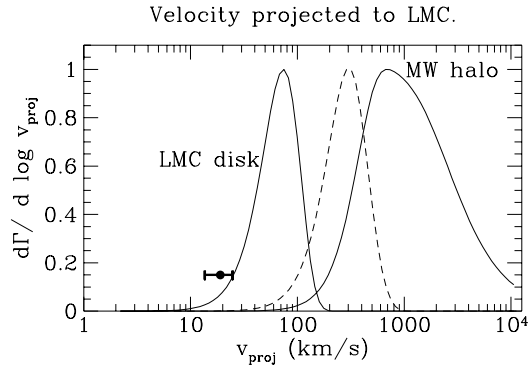}}
\end{center}
\vspace{-4.0cm}
\caption{The probability distributions of the $v_{\rm proj}$ are plotted
for lenses residing in the LMC disk and Milky Way halo. The dashed curve
represents the $v_{\rm proj}$ distribution for a super-massive LMC disk
with $\VEV{v^2} = 60\,$km/s. The $v_{\rm proj}$ value for the LMC binary
event is indicated with a horizontal error bar.}
\label{fig_vhat}
\end{figure}

Figure~\ref{fig_lmc_binary} shows the predicted $v_{\rm proj}$ distribution
for lenses in the LMC disk and Milky Way halo with the measured $v_{\rm proj}$
also indicated. Clearly, a lens in the LMC disk is prefered over a halo
lens. In fact, the $v_{\rm proj}$ value seems to be somewhat smaller than
expected for a lens in the LMC disk. This is, in fact, what one would
expect if the source was a short period binary rather than a single star.
Roughly speaking, we can expect binaries with periods of less than a few
hundred days to be separated by a distance much smaller than the Einstein
ring radius. From reference \cite{apt}, we see that about 20\% of late A star
binaries have periods of less than 100 days. Of these, about half will have
a companion that is at least 25\% as bright as the primary. Thus, a reasonable
a priori estimate of the fraction of caustic crossing events that will
have a detectable binary source character is about 10\%. If the source
star is a single star, then the measured $v_{\rm proj}$ value is only 
marginally consistent with lensing by an LMC object. For a binary source,
however, we would expect abnormally low values of $v_{\rm proj}$ for
lenses in the LMC, the Milky Way halo, or even the Milky Way disk. Thus,
we need to determine if the source star is actually a binary in order to
determine the location of the lens. It may be possible to accomplish this
by obtaining a single spectrum of the source star because late A stars appear
to have spectral peculiarities when they are in short period binary
systems \cite{apt}. This will be complicated by the fact that the source
star is superimposed on a brighter unlensed source, but analysis of the
image centroid motion during the lensing event indicates that the brighter
companion is actually 0.3" away from the lensed star. Thus, is
possible to get a spectrum with HST that can determine whether the
source star is likely to be a binary.

\begin{table}[htb]
\centering
\caption{ LMC Velocity Dispersion and 
               Self-Lensing Optical Depth} \label{tab_lmc}
\begin{tabular}{|ccc|} \hline
$v_{\rm rms}$ & $P(v_{\rm prop}=19\,{\rm km/s})$ & $\tau ($LMC-LMC$)$ \\ \hline
20 & 13\%  & $1.1\times 10^{-8}$ \\
25 & 7\%   & $1.8\times 10^{-8}$ \\
30 & 4\%   & $2.5\times 10^{-8}$ \\
40 & 1.8\% & $4.5\times 10^{-8}$ \\
50 & 0.9\% & $7.0\times 10^{-8}$ \\
60 & 0.5\% & $1.0\times 10^{-7}$ \\
75 & 0.3\% & $1.6\times 10^{-7}$ \\ \hline
\end{tabular}
\vspace{0.3cm}

The probability of obtaining the observed value of
$v_{\rm prop}$ and the LMC self-lensing optical depth are listed as a
function of the RMS line of sight velocity dispersion, $v_{\rm rms}$.
\end{table}

Now, let us assume that the source star is not a binary and consider the
implications if the lens is actually in the LMC. It might be tempting
to conclude that since the one lens that we can estimate the distance for 
appears to be in the LMC, perhaps LMC-LMC lensing might explain most or
even all of the lensing events. This would require that the microlensing
optical depth for LMC-LMC lensing be substantially in excess of the
value we have used in our microlensing background calculations. As Gould
\cite{gould} has shown, the self-lensing optical depth of 
a self-gravitating system is related to its velocity dispersion. For
a nearly face on disk like the LMC, the relationship is 
\begin{equation}
\tau = 2{\VEV{v^2}\over c^2} \sec^2 i = 2.52 {\VEV{v^2}\over c^2} ,
\label{eq_tauself}
\end{equation}
where we have used $i = 27^\circ$ as the inclination angle of the LMC disk. 

The $v_{\rm prop}$ distribution also depends on $v_{\rm rms}$. The solid
curve in Figure~\ref{fig_vhat} indicates the $v_{\rm prop}$ distribution
for $v_{\rm rms} = 25\,$km/s which is the measured value for CH stars
\cite{cow}. The dashed curve indicates the $v_{\rm prop}$ distribution for
$v_{\rm rms} = 60\,$km/s. Clearly, $v_{\rm prop} = 19\,$km/s is an unlikely
value for such a high line of sight velocity dispersion. Yet, the
self-lensing optical depth for $v_{\rm rms} = 60\,$km/s is still outside the
2-$\sigma$ confidence interval of the measured microlensing optical depth.
The relationship between $v_{\rm rms}$, the self-lensing optical depth and
the likelyhood of $v_{\rm prop} = 19\,$km/s is summarized in 
Table~\ref{tab_lmc}. $P(v_{\rm prop}=19\,{\rm km/s})$ is just
the relative likelyhood of $v_{\rm prop} = 19\,$km/s compared to the
most likely value for $v_{\rm prop}$. Only for $v_{\rm rms} = 75\,$km/s
is the self-lensing optical depth consistent with the 
2-$\sigma$ confidence interval on $\tau_{\rm meas}$, but this gives
a very small probability for the measured $v_{\rm prop}$ value. In this
case, the most reasonable explanation for the small $v_{\rm prop}$ value 
is the possibility of a binary source as discussed above. However, if the
source is a binary, then the argument favoring a lens in the LMC 
disappears. Thus, if the LMC self-lensing optical depth is significantly
smaller than $\tau_{\rm meas}$, then
there is some reason to suspect that the binary lens may reside in the LMC,
but if LMC self-lensing is responsible for most of
the total microlensing optical depth observed toward the LMC, then 
location of the LMC binary event cannot be determined from the caustic
crossing fit parameter. Clearly, the LMC binary event lends no weight to
the hypothesis that the LMC self-lensing optical depth is substantially
larger than $3\times 10^{-8}$ (the value used for our lensing background
calculations).

\end{document}